\documentclass[amsmath,amssymb,aps,prb,reprint,twocolumn,superscriptaddress,longbibliography,floatfix]{revtex4-2}
\usepackage{amsfonts}
\usepackage{graphicx}
\usepackage{xcolor}
\usepackage{units}  
\usepackage{float}
\usepackage{xspace}
\usepackage{subfigure}
\usepackage{caption}
\usepackage[colorlinks=true,allcolors=blue,breaklinks=true]{hyperref}
\usepackage{xurl}  
\usepackage{braket}
\usepackage{booktabs}
\usepackage{multirow}
\begin{document}
\title{Zero Indirect Band Gap in Non-Hermitian Systems}
	\author{S Rahul}
\affiliation{Department of Science and Humanities, PES University EC Campus, Bangalore 560100, India}

\author{Giandomenico Palumbo}
\affiliation{School of Theoretical Physics,
Dublin Institute for Advanced Studies, 
10 Burlington Road, Dublin 4, Ireland}

	\date{\today}
\begin{abstract}
    Zero indirect gaps in band models are typically viewed as unstable and achievable only through fine-tuning. Recent works, however, have revealed robust semimetallic phases in Hermitian systems where the indirect gap remains pinned at zero over a finite parameter range. Here, we extend this paradigm to non-Hermitian lattice models by studying a one-dimensional diamond-like system with gain and loss. We show that the zero indirect band gap in the real part of the spectrum remains
stable in the presence of non-Hermitian perturbations and identify the parameter regime in which this robustness persists. We find that the appearance of the zero indirect gap coincides with the suppression of the non-Hermitian skin effect. Our results reveal new connections between indirect gaps, exceptional points and non-Hermitian skin effect, opening avenues for experimental realizations.
\end{abstract}
\maketitle
\section{Introduction}
The distinction between direct and indirect band gaps plays a central role in determining the electronic and optical properties of condensed matter systems \cite{cooper2015indirect,malyi2020,10.1063/1.2177386}. While the direct band gap occurs at the same momentum points, the indirect band gap arises when the minimum of the conduction band and maximum of valence band occurs at different momenta. In this context, a zero indirect band gap is usually thought as unstable and induced by a mere fine tuning of the physical parameters of the systems.
However, it has been recently shown the existence of suitable one- \cite{palumbo2024topological} and two-dimensional Hermitian lattice models \cite{Palumbo2015,Juzeliunas,Pyrialakos2023,Wang2024}, in which the zero indirect band gap is robust and survives for a finite range of values of the physical parameters. This allows us to define of a novel family of semimetals in which bands do not touch each other differently from the more conventional semimetallic phases such as topological \cite{Armitage,Wan2011,Fang,Bradlyn2016,Zhu2020} and geometric semimetals \cite{Lin2024} that instead support Dirac-like and Weyl-like cones.
Importantly, one of these semimetallic phases with zero indirect band gap and non-trivial Chern number, dubbed two-dimensional Chern semimetal \cite{Palumbo2015}, has been experimentally realized in photonic crystals \cite{Chen2025}. Thus, it is then relevant at both theoretical and experimental level to figure out the behavior of these semimetallic phases in the presence of dissipation (gain and loss) and possibly show the existence of robust indirect band gaps in non-Hermitian phases.
We remind that in recent years, condensed and synthetic matter systems have been revisited from the perspective of non-Hermitian systems \cite{Ashida}, in which non-Hermitian Hamiltonians in open quantum systems emerge due to the presence of gain and loss. Here, the corresponding energy spectra can acquire complex values due to the nature of non-Hermiticity of the Hamiltonians. Non-Hermitian systems can exhibit distinctive features such as exceptional points (EPs) \cite{Heiss_2012,Alu,Budich2,Kawabata2019,Para2021}, the non-Hermitian skin effect (NHSE) \cite{Longhi2019,Li,Zhang31122022,PhysRevX.13.021007,Gohsrich_2025}, and non-Hermitian topology \cite{RevModPhys.93.015005, Esaki,Fu,Zhu2021,https://doi.org/10.1002/qute.202300225,martinezalvarez2018,
PhysRevX.8.031079,
PhysRevLett.123.066404,Borgnia,okuma}. 
In this framework, it is particularly intriguing to explore how zero indirect band gaps manifest within the complex band structures of non-Hermitian systems and how they influence the presence of EPs and the NHSE. Despite their significance, systematic studies at the intersection of non-Hermitian physics and indirect band gaps have been completely overlooked, providing a key motivation for the present work.\\
The main goal of this paper is to show the existence of robust zero indirect band gaps in non-Hermitian models.
Importantly, we refer to a zero indirect band gap exclusively in the real part of the energy spectrum. In fact, neither the imaginary part nor the modulus of the energy dispersion exhibits a zero indirect band gap.
For simplicity, we will focus on an one-dimensional system on a diamond-like lattice.
Along with the complex dispersions, we study the presence of EPs and the NHSE. By systematically varying the parameters of our non-Hermitian model, we will identify the ranges where the indirect band gap remains pinned at zero, thereby marking its stability.
We will show that in this regime the conventional NHSE is naturally suppressed due to the collapse of the point gap. While this suppression is expected from the symmetry of the couplings, it highlights a direct link between the spectral structure and the spatial distribution of eigenstates. Specifically, we will show that away from the condition $|\beta_1| = |\beta_2|$, the system exhibits a point gap and strong boundary localization characteristic of the NHSE. When the zero indirect band gap forms, the spectrum undergoes a transition to a line-gap-like structure, leading to partial delocalization of eigenstates and a significant reduction of the NHSE. This provides a clear illustration of how the band-structure properties govern the emergence and suppression of non-Hermitian skin effects in our system, differently from other mechanisms discussed in Refs \cite{PhysRevLett.127.256402,TEO20241667,qin2025,10.1063/5.0270893}.
Thus, our work paves the way for the exploration of new directions in non-Hermitian band theory.

\section{Model Hamiltonian}
\begin{figure}[H]
\centering
\includegraphics[width=0.7\columnwidth]{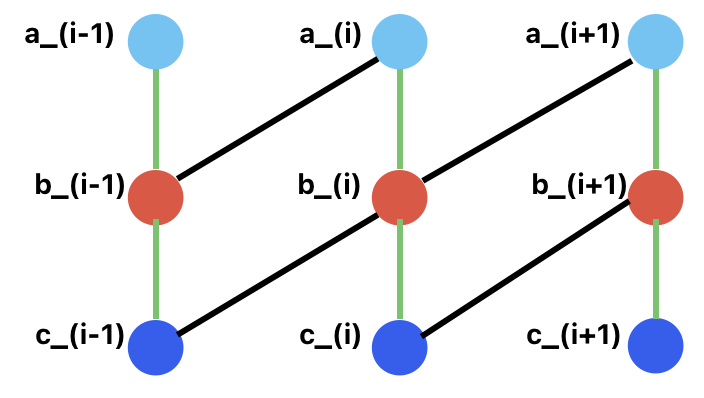}  
\caption{Diamond-like chain with three species of spinless fermions represented by light blue (a), red (b) and dark blue (c). Green lines correspond to inter species hopping while the dark lines corresponds to nearest neighbor hopping. }
\label{sch}
\end{figure}

We start considering a one dimensional system on a diamond-like chain with three sites per unit cell, which generalizes the Hermitian system introduced in Ref.~\cite{palumbo2024topological}. Schematic representation is shown in the fig.\ref{sch}.
The tight-binding Hamiltonian is written in terms of fermionic basis as,
 \begin{align}
  H &= \sum_i J\big(a^{\dagger}_{i+1} b_i + b^{\dagger}_{i+1} c_i + \text{h.c.}\big) 
      - h\big(a^{\dagger}_i b_i - b^{\dagger}_i c_i + \text{h.c.}\big) \nonumber \\
    &\quad + \beta_1 a^{\dagger}_i c_i + \beta_2 c^{\dagger}_i a_i,
\end{align}
where $J$ and $h$ are real parameters. $\beta_1$ and $\beta_2$ are generic complex numbers, which take into account gain and loss between $a$ and $c$ sites in the same unit cell (on-site dissipation).
The corresponding momentum space Hamiltonian is given by, 
\[
H(k) =
\begin{pmatrix}
0 & -h + J e^{-i k} & \beta_{1} \\
-h + J e^{i k} & 0 & -h + J e^{-i k} \\
\beta_{2} & -h + J e^{i k} & 0
\end{pmatrix}.
\]
Throughout the manuscript $J$ and $h$ are set to 1 for simplicity. Despite this choice, it is possible to show that the zero indirect gap persists even for other values of the hopping parameters.
It is important to note that for generic values of the parameters $\beta_1 \neq \beta_2$, the Bloch Hamiltonian is non-Hermitian and its spectrum is determined by a cubic characteristic equation. Although the eigenvalues can be obtained algebraically using Cardano’s formulas, the resulting expressions involve nested square and cube roots of complex, momentum-dependent functions and are therefore intrinsically multi-valued functions of the crystal momentum $k$. The branching structure of these solutions is controlled by the discriminant of the cubic equation: points in parameter space and momentum where the discriminant vanishes correspond to exceptional points, at which eigenvalues and eigenvectors coalesce. As a result, the spectrum cannot be represented as globally single-valued analytic band dispersions across the Brillouin zone, and no unique band labeling exists. This analytic obstruction is absent in the special case $\beta_1=\beta_2=0$, where the model reduces to a Hermitian form and the spectrum admits a simple, single-valued analytic expression, given by
a perfect flat middle band and two linear dispersive bands, such that all three bands touch each other at a Dirac-like point at $k=0$.
\begin{figure}[H]
\centering

{
    \includegraphics[width=0.8\columnwidth]{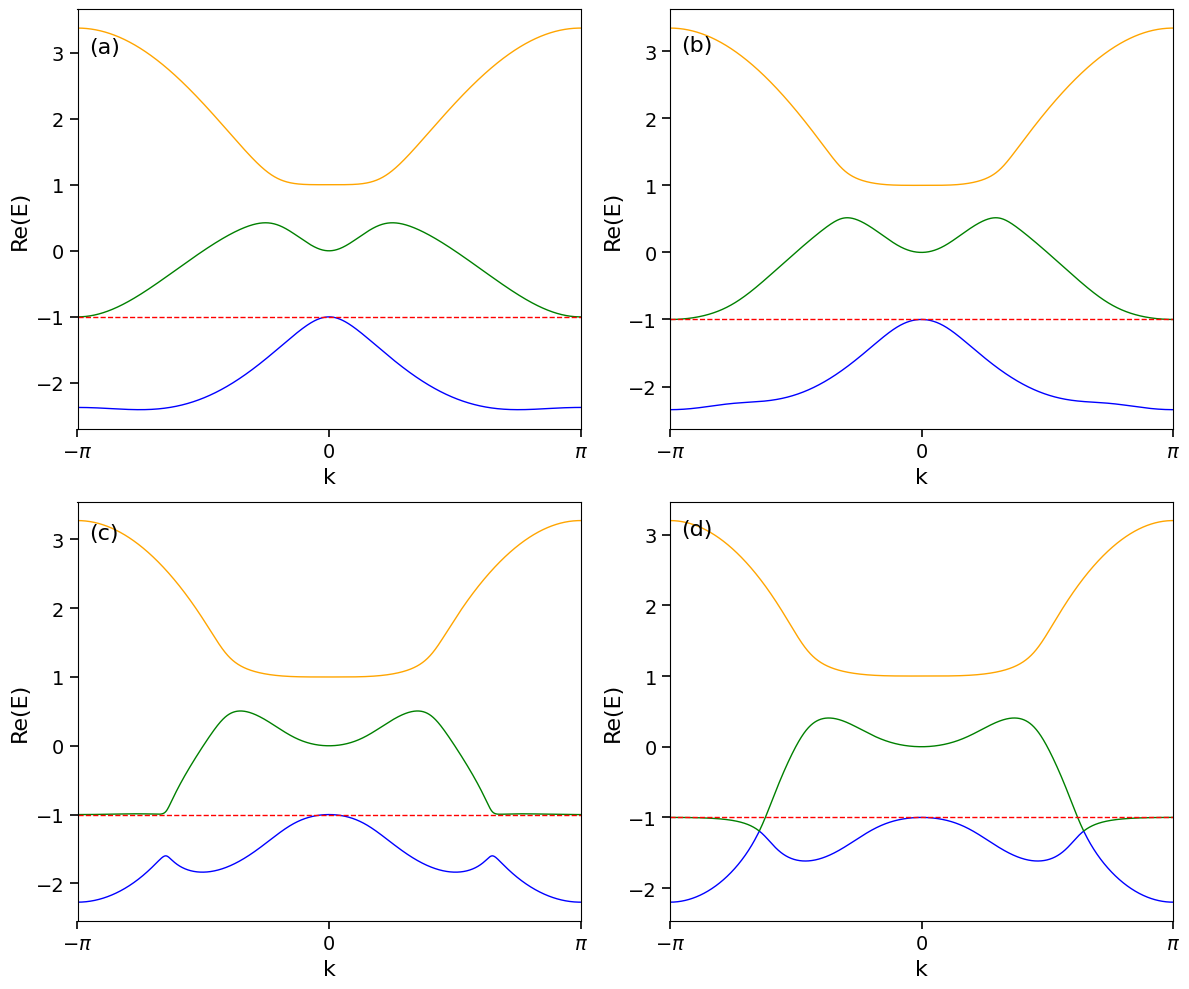}
    \label{gen1}
}

\vspace{0.5em}

{
    \includegraphics[width=0.8\columnwidth]{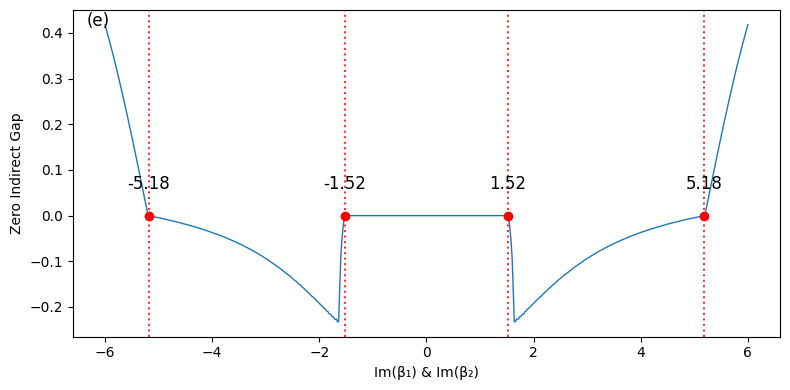}
    \label{gen11}
}

\caption{(a)-(d) Real component of the dispersion spectrum plotted as a function of k. Panels (a)-(d) correspond to distinct set of imaginary part of $Im(\beta_1) = Im(\beta_2) = 0.2, 0.8, 1.52, 2.0$ respectively. (e) Zero indirect gap plotted with respect to the $Im(\beta_1) = Im(\beta_2)$ for fixed value of $Re(\beta_1) = Re(\beta_2)$=1.} 
\label{combined}
\end{figure}
\begin{figure}[H]
\centering
\includegraphics[width=0.95\columnwidth]{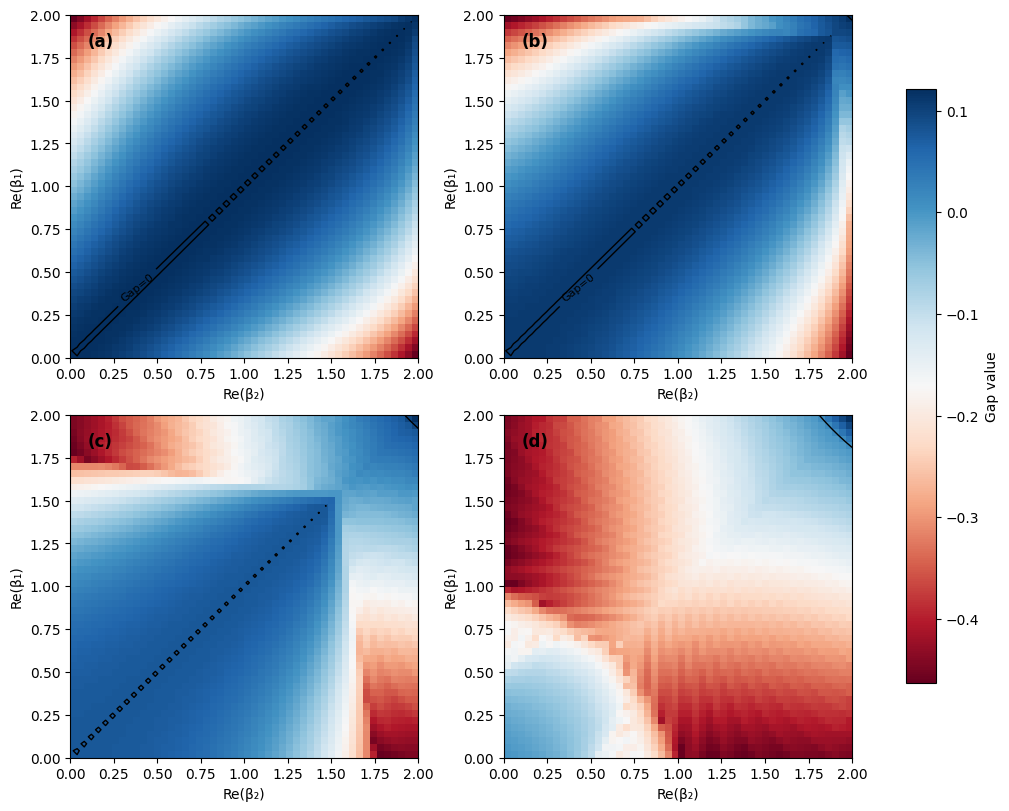} 
\caption{Parameter space plotted with respect to $Re(\beta_1)$ and $Re(\beta_2)$ for different values of imaginary components i.e., (a)-(d) $Im(\beta_1) = Im(\beta_2) = 0.2, 0.6, 1.2, 1.8$ respectively.}
\label{gen2}
\end{figure}
The momentum–space dispersion, shown in Fig.~\ref{combined}(a)-(c), reveals the emergence of a zero indirect band gap. This feature persists across different values of the imaginary components of $\beta_{1}$ and $\beta_{2}$, while keeping their real parts fixed at $ Re(\beta_{1}) = Re(\beta_{2}) = 1$. 
A horizontal red dashed line serves as a marker to indicate the presence of zero indirect band gap.  
From Fig.~\ref{combined}(d), it is evident that as the imaginary parts of $\beta_{1}$ and $\beta_{2}$ increase, the system undergoes a qualitative change in its band structure i.e., the zero indirect band gap gradually disappears, giving rise to a direct gap closure.\\ 
This signifies that the enhancement in the strength of non-Hermitian coupling, destabilizes the zero indirect band gap. The range of $Im(\beta_1) = Im(\beta_2)\equiv Im(\beta)$ for which the zero indirect band gap exist is presented in Fig.\ref{combined}(e) and is given by $-1.52<Im(\beta)<1.52$. The existence of this finite range demonstrates that the zero indirect bang gap is indeed robust.   
To further investigate this regime, we map out the parameter space in terms of the real components of $\beta_1$ and $\beta_2$.
In the Fig.\ref{gen2}, we illustrate how a zero indirect band gap emerges within certain regions of the parameter space. This regime is analyzed for various values of $Im(\beta)$ indicating that it exists under the condition $Re(\beta_1) = Re(\beta_2)$.
\section{Non-Hermitian Skin effect}
\begin{figure}[H]
\centering
\includegraphics[width=1.0\columnwidth]{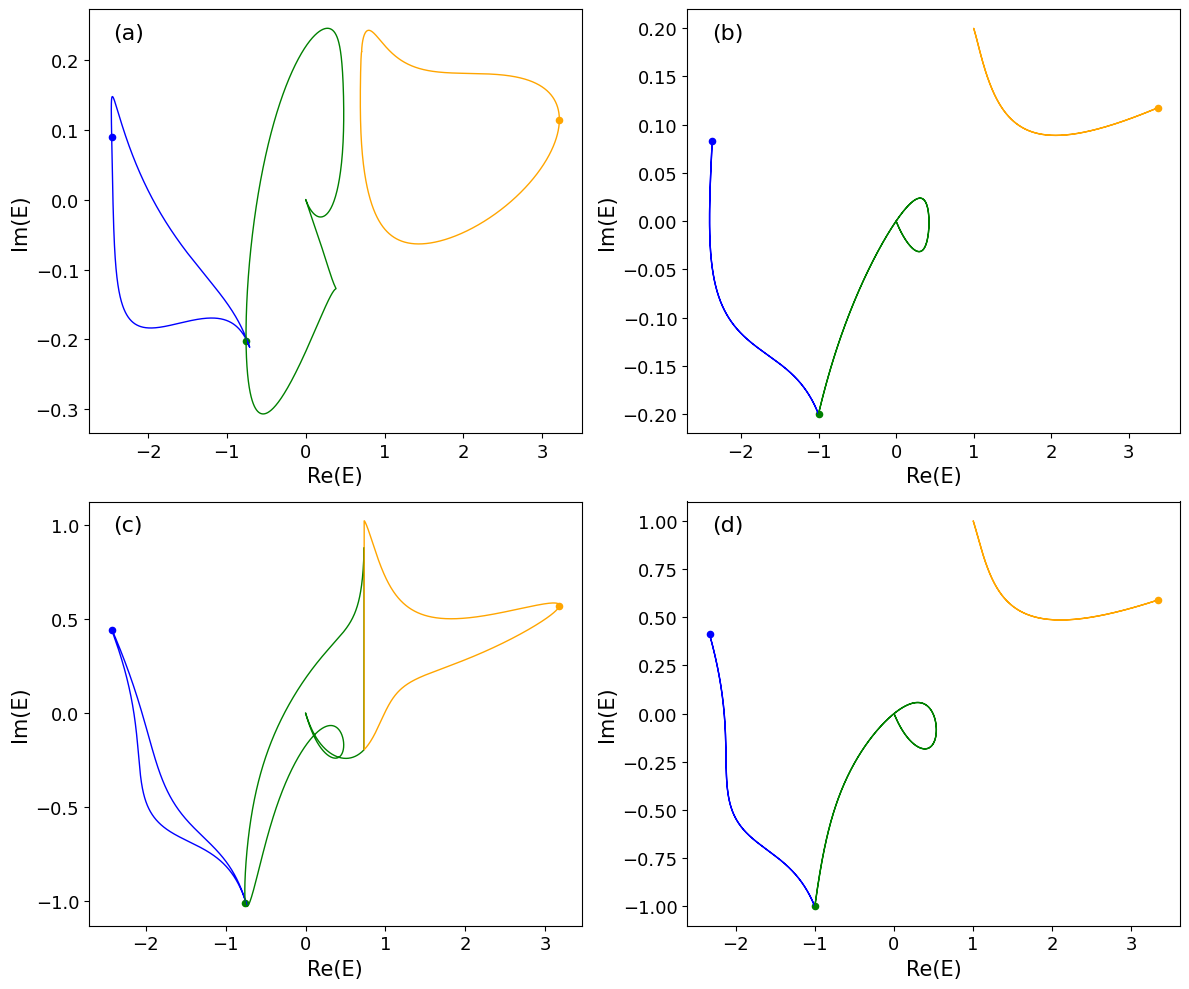}   
\caption{The complex energy spectra obtained by varying the imaginary parts of the non-Hermitian parameters
$\beta_1$ and $\beta_2$, while keeping their imaginary parts fixed at
$\mathrm{Im}(\beta_1)=\mathrm{Im}(\beta_2)=0.2,1.0$. } 
\label{gap}
\end{figure}

Before reviewing the NHSE in our model, it is useful to briefly discuss the central role of NHSE in non-Hermitian systems. This peculiar effect refers to the significant number of bulk eigenstates accumulate at the boundary under open boundary conditions \cite{Zhang31122022,PhysRevX.13.021007,Gohsrich_2025}. Physically the NHSE arises due to the asymmetric couplings or gain/loss \cite{PhysRevB.110.155144} in the system. The existence of non-Hermiticity in both quantum materials and synthetic matter systems makes the NHSE very relevant from the perspective of experimental realizations \cite{Wang_2021,longhi2015,PhysRevB.92.094204}. 
Here, to characterize this effect induced by non-Hermitian couplings $\beta_1$ and $\beta_2$ in our model, we compute the average inverse participation ratio (IPR) of the eigenstates across the parameter space spanned by Re($\beta_1$) and Re($\beta_2$). 
The IPR is a quantitative measure of spatial localization of eigenstates. For a normalized eigenstate $\psi_n$ defined over lattice sites L, the IPR is given by, 
\begin{equation}
\mathrm{IPR} = \sum_{j=1}^{L} \left| \psi_n(j) \right|^4 ,
\label{eq:ipr_def}
\end{equation}
where $\psi_n(j)$ denotes the amplitude of the $n$-th eigenstate at site $j$.\\ 
The results reveal a clear suppression of NHSE along the diagonal line Re($\beta_1$) = Re($\beta_2$), where the non-Hermitian coupling is effectively symmetric and hence the asymmetry vanishes. In this regime the eigenstates are more delocalized, leading to smaller IPR values. In contrast, away from the diagonal, the asymmetry between Re($\beta_1$) and Re($\beta_2$) becomes significant, and the eigenstates exhibit strong localization reflected in the higher IPR values. 
\begin{figure}[H]
\centering
\includegraphics[width=1.0\columnwidth]{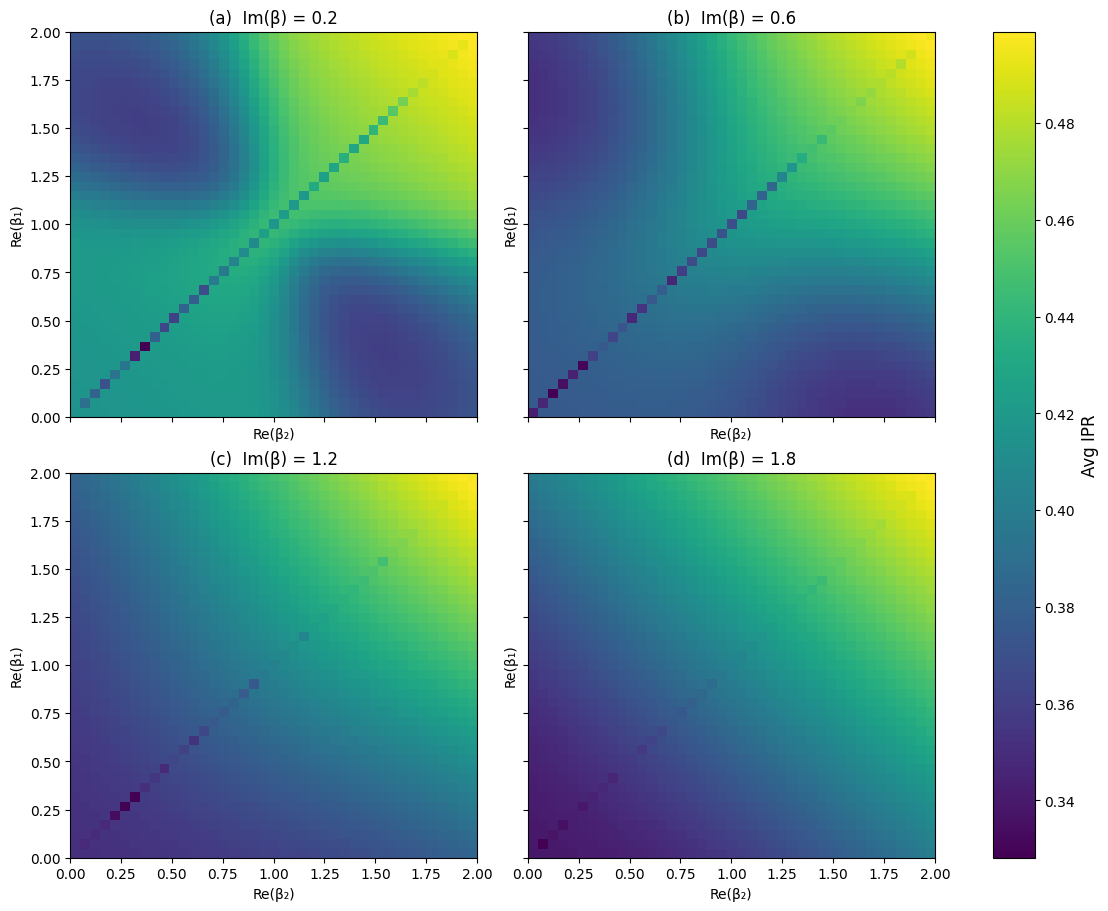} 
\caption{Average IPR of eigenstates as a function of $\mathrm{Re}(\beta_1)$ and $\mathrm{Re}(\beta_2)$ for different values of the imaginary component $\mathrm{Im}(\beta)$ (a) $0.2$, (b) $0.6$, (c) $1.2$, and (d) $1.8$.}
\label{gen7}
\end{figure}
The clear suppression of the NHSE that coincides with the appearance of the zero indirect bandgap can further justified by studying the behavior of all three bands in terms of point gap and line gap topology. Here in the fig.\ref{gap}, we present the imaginary vs real eigenvalue plot to understand the nature of all three bands for different imaginary $\beta_1$ and $\beta_2$ values.\\ 
Figure~\ref{gap} shows the complex energy spectra plotted as $\mathrm{Im}(E)$ versus $\mathrm{Re}(E)$ for different choices of the non-Hermitian parameters $(\beta_1,\beta_2)$. Each panel corresponds to a distinct regime of non-reciprocity and gain and loss strength. In panels (a) and (c), where $\mathrm{Re}(\beta_1)\neq \mathrm{Re}(\beta_2)$, the complex bands form extended loops that enclose a finite area in the complex-energy plane. Such spectral loops indicate the presence of a \emph{point gap} rather than a line gap which is known to be directly associated with the emergence of the NHSE.\\
This behavior is consistent with the phase diagram fig.\ref{gen7}, where the average IPR is plotted in the $\mathrm{Re}(\beta_1)$--$\mathrm{Re}(\beta_2)$ plane for different values of $\mathrm{Im}(\beta)$. Away from the diagonal line $\mathrm{Re}(\beta_1)=\mathrm{Re}(\beta_2)$, the IPR remains relatively large, signaling strong boundary localization of eigenstates and hence a pronounced skin effect.

In contrast, panels (b) and (d) of Fig.~\ref{gap}, corresponding to $\mathrm{Re}(\beta_1)=\mathrm{Re}(\beta_2)$, exhibit a qualitatively different structure. Here the complex bands are significantly compressed and forms a arc. This indicates the collapse of the point gap and the emergence of an effective \emph{line gap-like} structure in the real part of the spectrum. 
In this regime, the zero indirect bandgap of the real part of the spectrum is present (refer fig.\ref{gen2}) and as a result, the band structure goes from point gap to line-like gap, presented in the fig.\ref{gap}. 
Physically, the presence of a zero indirect band gap constrains the real parts of different bands to touch at distinct momenta, which inhibits the formation of arcs in the complex energy plane. As a result, the non-reciprocal accumulation of eigenstates is suppressed, leading to the lowering of the NHSE despite the system remaining non-Hermitian.

\section{Exceptional Points}

In this section, we investigate the appearance of exceptional points (EPs) in the direct-gap regime, which arise due to the non-Hermitian coupling terms in our model. EPs occur in the parameter space where all three bands coalesce at isolated points. It is important to emphasize that EPs are absent in the zero indirect-gap regime. Therefore, to characterize the behavior away from the zero-gap regime, we compute the condition number ${\rm cond(V)}$ \cite{g1cw-tk7f,PhysRevResearch.6.013044,PhysRevResearch.4.023121}, which provides a quantitative measure of the proximity to EPs, by identifying the non-orthogonality of eigenvectors, which can become nearly linearly dependent.\\
The right eigenvectors of the Hamiltonian 
$H(\mathbf{k})$ are obtained by solving
\begin{align}
    H(\mathbf{k}) \, |R_n(\mathbf{k})\rangle = E_n(\mathbf{k}) \, |R_n(\mathbf{k})\rangle ,
\end{align}
where $E_n(\mathbf{k})$ are (generally complex) eigenvalues. 
Writing these right eigenvectors as columns of a matrix $V(k)$, 
\begin{align}
    V(\mathbf{k}) = \big( |R_1(\mathbf{k})\rangle, |R_2(\mathbf{k})\rangle, \dots, |R_N(\mathbf{k})\rangle \big),
\end{align}
the condition number of $V(k)$ is defined as
\begin{align}
    \mathrm{cond}\big(V) 
    = \| V(\mathbf{k}) \| \, \| V(\mathbf{k})^{-1} \|.
\end{align}
\begin{figure}[H]
\centering
\includegraphics[width=1.0\columnwidth]{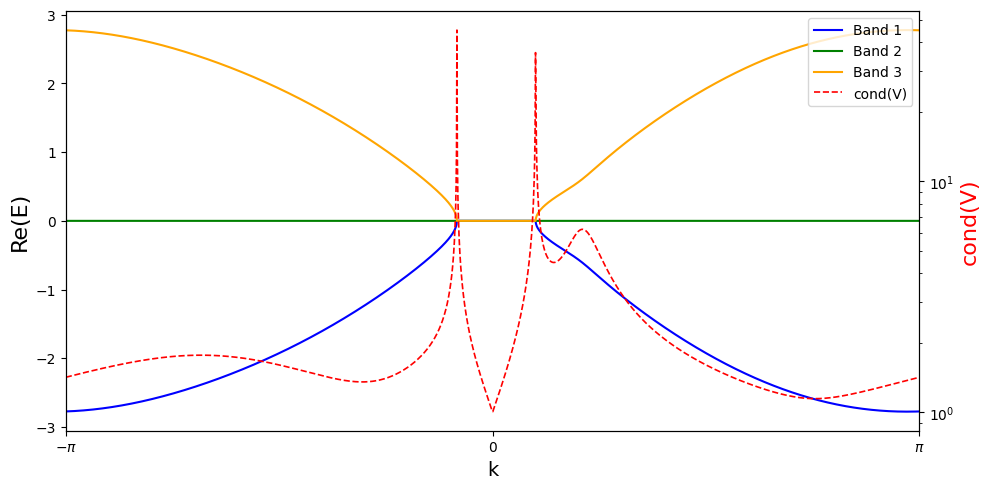}   
\caption{Real component of the dispersion spectrum plotted with respect to k. The red dotted line corresponds to the profile of ${\rm cond(V)}$ plotted over k for the values $\beta_1 = -0.5+0.5i$ and $\beta_2 = 0.5+0.5i$.}
\label{gen5}
\end{figure}
At EPs, not only two or more eigenvalues coalesce, 
but their associated eigenvectors also become linearly dependent. 
In this case, the matrix $V(\mathbf{k})$ becomes nearly singular, and hence ideally the condition number diverges to infinite. 
In Fig.~\ref{gen5}, we present the behavior of the condition number 
${\rm cond(V)}$ superimposed on the energy dispersion spectrum. 
The figure clearly illustrates that the sharp divergence of 
${\rm cond(V)}$ occurs precisely at the points in momentum space where the eigenvalue bands coalesce. At these points, not only the eigenvalue bands but also the corresponding eigenvectors coalesce. This feature is indicated by the divergence of ${\rm cond(V)}$. 
\begin{figure}[H]
\centering
\includegraphics[width=0.8\columnwidth]{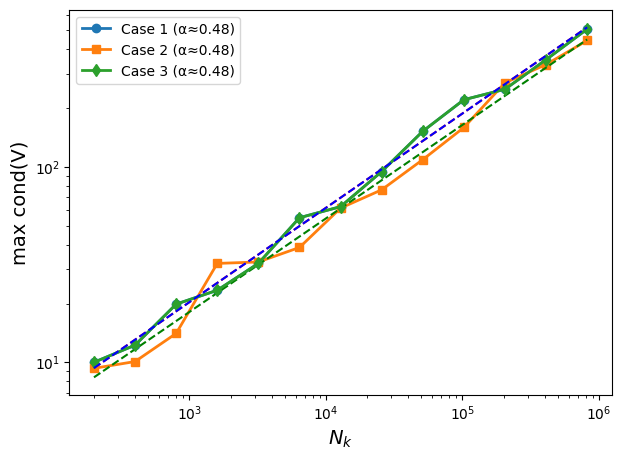}   
\caption{Scaling behavior of the maximum condition number of the matrix V(k) as a function of the number of momentum points $N_k$ for different parameter choices of $\beta_1$ and $\beta_2$, namely $\beta_1 = \{-0.5,1.0\}, \; \beta_2 = \{0.5,1.0\}$; 
         $\beta_1 = \{0.25,1.0\}, \; \beta_2 = \{-0.25,1.0\}$; and 
         $\beta_1 = \{0.5,1.0\}, \; \beta_2 = \{-0.5,1.0\}$ respectively.
} 
\label{gen6}
\end{figure}
This divergence serves as a marker of EPs, highlighting the 
loss of linear independence of the eigenvectors and the consequent ill-conditioning of the eigenvector matrix $V(k)$. 
 Moreover to understand if the maximum value of ${\rm cond(V)}$ scales with the resolution of k, we do a systematic scaling analysis which is presented in Fig.\ref{gen6}.
 Across all three cases, the scaling law follows an approximate power-law behavior of the form, 
 \begin{equation}
    \text{maximum cond}(V) \propto N_{k}^{\alpha},
\end{equation}
with a fitted exponent $\alpha=0.48$. The consistency of this exponent across different parameter regimes highlights the robustness of the scaling behavior. 
This robustness not only provides a reliable quantitative signature for the presence of EPs but also suggests a universal mechanism governing the sensitivity of eigenvectors near non-Hermitian degeneracies in our model. 

\section{Conclusion}
In this work, we have investigated the emergence of a robust zero indirect band gap of the real part of the spectrum related to a non-Hermitian diamond-like chain with generic complex intra-cell couplings. By analyzing the momentum space dispersion spectrum, we have identified the parameter regime where the zero indirect band gap persists. The non-Hermitian character of the model Hamiltonian leads to rich phenomena such as the emergence of EPs, NHSE. Our results reveal that the appearance of zero indirect band gap coincides with the suppression of the NSHE. 
It would be relevant to check the weakening of the NHSE also in higher-dimensional gapless systems with zero indirect gaps. We have also computed condition number of the eigenvector matrix to identify the EPs and have obtained a scaling law that proves the divergence of condition number at the EPs. We leave this important point to future work. This interplay between zero indirect band gap and the non-Hermitian features provides new insights into spectral properties of non-Hermitian systems and supports in engineer novel lattice systems. 
Finally, we note that while our study is primarily theoretical, the proposed non-Hermitian Hamiltonian on a diamond-like chain could be implemented in existing experimental platforms, including photonic lattices \cite{Mukherjee}, cold atoms in optical lattices \cite{Hyrkas}, and engineered solid-state systems \cite{Huda}. In particular, controllable gain and loss, as well as asymmetric hopping, could be introduced using Floquet engineering \cite{palumbo2024topological} or other lattice-synthesis techniques. 
As a future direction, we plan to investigate non-Hermitian phases in topological semimetals exhibiting robust zero indirect gaps in two \cite{Palumbo2015} and higher dimensions, aiming to uncover novel spectral and transport phenomena arising from the interplay between topology and dissipation.

\bibliography{indirect}




\end{document}